\documentstyle[aps]{revtex}
\begin{document}
\draft
\title{Black hole entropy: certain quantum features}
\author{Parthasarathi Majumdar}
\address{The Institute of Mathematical Sciences, Chennai 600 113,
India\\Email:partha@imsc.ernet.in}
\maketitle
\begin{abstract}
The issue of black hole entropy is reexamined within a finite lattice framework
along the lines of Wheeler, 't Hooft and Susskind, with an additional criterion
to identify physical horizon states contributing to the entropy. As a
consequence, the degeneracy of physical states is lower than that attributed
normally to black holes. This results in corrections to the Bekenstein-Hawking
area law that are logarithmic in the horizon area.  Implications for the
holographic entropy bound on bounded spaces are discussed. Theoretical
underpinnings of the criterion imposed on the states, based on the `quantum
geometry' formulation of quantum gravity, are briefly explained.
\end{abstract}

\section{Introduction}

The notion of black hole entropy, introduced by Bekenstein \cite{bek1} on the
basis of the laws of black hole mechanics \cite{haw1} and gleaning insights from
(classical) information theory, is one of the most profound in black hole
physics. The issue of which microstates contribute to the entropy has remained a
challenging one, since these states should ostensibly appear in a quantum theory
that includes gravitation. In the absence (even now) of a complete quantum
theory of gravitation, a measure of the entropy was suggested on semiclassical
grounds \cite{bek1}, \cite{haw2} as being equal to a quarter of the horizon area
of the black hole. Deeply engaging insights into this unexpected dependence of
entropy on surface area (rather than volume) have appeared early in the last
decade. The `It from bit' picture of Wheeler \cite{whe} contains the germ of
these insights which have subsequently been substantially refined \cite{thf}-
\cite{suss}, leading to the bold proposal of the principle of Holography, as
applied to quantum gravity. In this paper, some of these insights are reexamined
from a somewhat different standpoint. The departure from standard lore appears
to bring into the fold the first truly `quantum gravity' aspects beyond the
semiclassical Bekenstein-Hawking Area Law (BHAL). 

The paper is organized as follows: in Section 2, we survey the basic ingredients
of the `It from bit' picture; in Section 3, we discuss the criterion we impose on
the space of states to identify the physical Hilbert space of horizon states; the
dimensionality of this physical Hilbert space yields the entropy of the black
hole which has a $\log (area)$ correction over and above the BHAL. This is argued
to lead to an upper bound on the entropy of all bounded three-spaces (where the
boundary is $S^2$) via the holographic principle. In Section 4, we {\it derive}
our physical subspace criterion of the earlier section on the basis of the
microscopic theory of Quantum Geometry. Our concluding remarks are presented in
Section 5.

\section{`It from bit'}

Consider a two dimensional finite `floating lattice' with plaquettes
approximately the size of a Planck area ($\sim l_P^2$) covering the
spherical horizon of an eternal non-rotating four dimensional black hole.
The black hole is assumed to be macroscopic\footnote{Planck-size or
primordial black holes fall outside the purview of this work} in that the
classical area of its horizon $A_S/l_P^2 \gg 1$.  Assume that binary
variables (`bits', `Boolean variables' or `pixels') are distributed
randomly on this lattice. Typically, these could be elementary spin 1/2
variables or doublets of an $SU(2)$ group. Assume also that the size of
the lattice is characterized by a finite large even integer $p$.  Clearly,
the Hilbert space of quantum states defined by these spin 1/2 variables
has a dimensionality ${\cal N}(p) = 2^p$. It follows that the number of
degrees of freedom characterizing the horizon is given by $N \equiv \log
{\cal N}(p) = p \log 2$. Given the relation between entropy and number of
degrees of freedom, the former is also proportional to the size $p$ of the
lattice. By assumption, $p \gg 1$; in the limit of very large $p$, the
lattice can be taken to approximate the macroscopic horizon of the black
hole. One would then expect that the classical horizon area $A_S$ would
satisfy $A_S/l_P^2 = \xi~p$ where $\xi = O(1)$. For a choice $\xi = 4 \log
2$, one obtains for the entropy $S_{bh} \equiv N = A_S/4l_P^2$ which is
the famous BHAL.

The generality of the above scenario makes it appealing vis-a-vis
a quantum theory of black holes in particular and of quantum gravity in
general.  There is however one crucial aspect of any quantum approach to
black hole physics which seems to have been missed in the above, -- the
aspect of symmetry.  Indeed, the mere random distribution of spin 1/2
(binary) variables on the lattice which approximates the black hole
horizon, without regard to possible symmetries, possibly leads to a far bigger
space of states than the {\it physical} Hilbert space, and hence to an
overcounting of the number of the degrees of freedom, i.e., a larger
entropy.

\section {The physical Hilbert (sub)space}

\subsection{A natural symmetry criterion}

But what is the most plausible symmetry that one can impose on states so as to
identify the physical subspace ? Recall that the elementary variables are binary
or spin 1/2 variables which can be considered to be in the fundamental doublet
representation of an $SU(2)$ Lie Algebra. On very general grounds then, the most
natural symmetry of the physical subspace must be this $SU(2)$. One is thus led
to a symmetry criterion which identifies the physical Hilbert space ${\cal H}_S$
of horizon states contributing to black hole entropy: {\it ${\cal H}_S$ consists
of states, composed of elementary $SU(2)$ doublets, which are $SU(2)$ singlets}. 
Observe that this criterion has no allusions whatsoever to any specific proposal
for a quantum theory of gravitation. Nor does it involve any gauge redundancies
(or any other infinite dimensional symmetry like conformal invariance) at this
point. It is the most natural choice for the symmetry of physical horizon states
simply because in the `It from bit' picture, the basic variables are spin 1/2
variables. Were they of higher multiplicity than 2, the Lie Algebra of
symmetries might have been likewise different. Later on we shall show however
that this symmetry arises very naturally in the Quantum Geometry approach to
black hole physics. It will emerge from that approach that horizon states of
large macroscopic black holes are best described in terms of spin 1/2 variables
at the punctures of a punctured two-sphere which represents (a spatial slice of)
the event horizon. 

\subsection{Dimensionality of ${\cal H}_S$}

The criterion of $SU(2)$ invariance leads to a simple way of counting the
dimensionality of the physical Hilbert space, as has already been shown
\cite{dkm}. For $p$ variables, this number is given by 
\begin{equation}
dim{\cal H}_S \equiv {\cal N}(p)~=~\left( \begin{array}{c}
                         p \\ p/2
                        \end{array} \right)
                 ~ - ~\left( \begin{array}{c}
                         p \\ (p/2-1)
                         \end{array} \right)  ~\label{enpo}
\end{equation}
There is a simple intuitive way to understand the result embodied in
(\ref{enpo}). This formula counts the number of ways of making
$SU(2)$ singlets from $p$ spin $1/2$ representations. The first term   
corresponds to the number of states with net $J_3$ quantum number $m=0$
constructed by placing $m=\pm 1/2$ on the punctures.  However, this
term by itself {\it overcounts} the number of $SU(2)$ singlet states,
because even non-singlet states (with net integral spin, for $p$ is
an even integer) have a net $m=0$ sector. Beside having a sector with
total $m=0$, states with net integer spin have, of course, a sector with
overall $m=\pm 1$ as well. The second term basically eliminates these
non-singlet states with $m=0$, by counting the number of states
with net $m=\pm 1$ constructed from $m=\pm 1/2$ on the $p$
sites. The difference then is the net number of $SU(2)$ singlet
states that represents the dimensionality of ${\cal H}_S$.

It may be pointed out that the first term in (\ref{enpo}) also has another
interpretation. It counts the number of ways binary variables corresponding to
spin-up and spin-down can be placed on the sites to yield a vanishing total spin.
Alternatively, one can think of the binary variables as unit positive and negative
$U(1)$ charges; the first term in (\ref{enpo}) then corresponds to the
dimensionality of the Hilbert space of $U(1)~invariant$ states. As already shown in
\cite{dkm}, this corresponds to a {\it binomial} rather than a random distribution
of binary variables. 

\subsection{Large $p$ approximation and black hole entropy}

In the limit of very large $p$, one can evaluate the factorials in (\ref{enpo})
using the Stirling approximation. One obtains
\begin{equation}
{\cal N}(p)~\approx~ {2^p \over p^{\frac32} }~. \label{plar}
\end{equation}
Clearly, the dimensionality of the physical Hilbert space is smaller than what one
had earlier, as would be an obvious consequence of imposing $SU(2)$ symmetry. Using
the relation between $p$ and the classical horizon area $A_S$ discussed in the last
section, with the constant $\xi$ chosen to take the same value as in that section,
(\ref{plar}) can be shown \cite{dkm} to lead to the following formula for black
hole entropy,
\begin{equation}
S_{bh}~\equiv~\log {\cal N}(p)~\approx~  {A_S\over{4l_P^2}} ~-~{3\over
2}~\log \left({A_S\over{4l_P^2}} \right)~+~ const.~+~O(A_S^{-1}).
\label{main}  
\end{equation}

The general nature of the assumptions underlying the derivation of eq. 
(\ref{main}) can hardly be overemphasized. Nowhere has any particular aspect of
a {\it microscopic} theory of quantum gravity been used. Nor did we need to
appeal to any specific infinite dimensional symmetry of the classical horizon
geometry, and formulas pertaining to such an invariance. In spite of this, the
physical subspace criterion discussed earlier in this section, leads
unequivocally to the BHAL, and a correction term logarithmic in the horizon
area, with the coefficient -3/2. Of course, the constant $\xi$ was chosen to
have a fixed value, in order to fix the normalization of the BHAL. Observe
though that the coefficient of the $\log (area)$ correction is unaffected by
the choice of $\xi$. There are further sub-leading corrections that are
constant and inversely proportional to powers of horizon area, as an expansion
for large areas should entail. 

This result is of course not new; it has been derived earlier \cite{km2} within
the context of quantum geometry, where the correct normalization of the BHAL
dictates a choice of the Barbero-Immirzi parameter. Once again, the coefficient
of the logarithmic correction is independent of this choice. Rather compelling
arguments have been presented (from a perspective different from ours) \cite{car}
that the correction could be of a `universal' character. Our discussion above
would lend credence to such an inference on more general grounds. But, perhaps
of similar significance is the feature that {\it kinematically}, quantum horizon
states of large macroscopic non-rotating black holes appear to have a rather
simple description in terms of $SU(2)$ singlet states of a (two dimensional) 
lattice of spin 1/2 variables. It turns out that such a description is not
restricted to static black hole horizon states, but in fact, describe Isolated
Horizons \cite{ash}, \cite{ash2} of macroscopic size as well. 

\subsection{Holography and the entropy bound}

Having identified the kinematical quantum states characterising a black hole
horizon, the question that immediately comes to mind is whether there are other
states that describe black hole physics. Although the `It from bit' picture tends
to imply that the entire information lies with the horizon states, this has been
more sharply articulated in the so-called Holographic Principle \cite{thf}.
According to this principle, the horizon states exhaust the Hilbert space of a
black hole, encoding the entire information of gravitationally collapsed matter in
terms of macroscopic observables like the horizon area. The entropy of a black
hole can then be taken to represent the {\it maximal} possible entropy of a
spacetime whose spatial slice has a boundary that coincides with the intersection
of this spatial slice with the horizon. Now, it can be shown \cite{dkm} that eq.
(\ref{main}) actually translates into a bound on black hole entropy, given
by
\begin{equation}
S_{max}~=~ln \left( {\exp{S_{BH}} \over S_{BH}^{3/2} } \right)~
\label{newb1}
\end{equation}

Thus, it follows that {\it all} 3-spaces with boundary have an entropy bounded
from above by (\ref{newb1}). That this is extremely plausible follows from the
following argument, based on {\it reductio ad absurdum} \cite{smo}:  we assume,
for simplicity that the spatial slice of the boundary of an asymptotically flat
spacetime has the topology of a 2-sphere on which is induced a spherically
symmetric 2-metric. Let this spacetime contain an object whose entropy exceeds
the entropy bound given in eq. (\ref{newb1}). Certainly, such a spacetime
cannot have a black hole horizon as a boundary, since then, its entropy would
have been subject to (\ref{newb1}). But, in that case, its energy should be
less than that of a black hole which has the 2-sphere as its horizon.  Let us
now add energy to the system, so that it does transform adiabatically into a
black hole with the said horizon, but without affecting the entropy of the
exterior. But we have already seen above that a black hole with such a horizon
must obey the bound (\ref{newb1}); it follows that the starting assumption of
the system having an entropy exceeding (\ref{newb1}) must be incorrect. Thus,
we have indeed obtained an upper bound on the entropy of a large class of
spacetimes.  Notice that this bound {\it tightens} the semiclassical Bekenstein
bound \cite{bek2}, which is of course expected because of its quantum
kinematical underpinning. We now turn to an expos\'e of this underlying
structure within the framework of Quantum Geometry. 

\section{Entropy from Quantum Geometry}

The presentation in this section closely follows ref. \cite{dkm}; we consider
generic 3+1 dimensional Isolated Horizons without rotation, on which one assumes
an appropriate class of boundary conditions \cite{ash}.  These boundary conditions
require that the gravitational action be augmented by the action of a Chern-Simons
theory living on the isolated horizon.  Boundary states of the Chern-Simons theory
contribute to the entropy. These states correspond to conformal blocks of the
two-dimensional Wess-Zumino model that lives on the spatial slice of the horizon,
which is a 2-sphere of area $A_H$.  The dimensionality of the boundary Hilbert
space has been calculated thus \cite{km2}, \cite{km} by counting the number of
conformal blocks of two-dimensional $SU(2)_k$ Wess-Zumino model, for arbitrary
level $k$ and number of punctures $p$ on the 2-sphere. We shall show, from the
formula for the number of conformal blocks specialized to macroscopic black holes
characterized by large $k$ and $p$ \cite{km2}, that eq. (\ref{main}) ensues. 

Let us start with the formula for the number of conformal blocks of
two-dimensional $SU(2)_k$ Wess-Zumino model that lives on the punctured
2-sphere. For a set of punctures ${\cal P}$ with spins $ \{j_1, j_2, \dots
j_p \} $ at punctures $\{ 1,2, \dots, p \}$, this number is given by
\cite{km}
\begin{equation}
N^{\cal P}~=~{2 \over {k+2}}~\sum_{r=0}^{ k/2}~{
{\prod_{l=1}^p sin \left( {{(2j_l+1)(2r+1) \pi}\over k+2} \right) }
\over
{\left[ sin \left( {(2r+1)  \pi \over k+2} \right)\right]^{p-2} }} ~.
\label{enpi}
\end{equation}
Observe now that Eq. (\ref{enpi}) can be rewritten as a multiple sum,
\begin{equation}
N^{\cal P}~=~\left ( 2 \over {k+2} \right) ~\sum_{l=1}^{k+1} sin^2
\theta_l~\sum_{m_1 =
-j_1}^{j_1} \cdots \sum_{m_p=-j_p}^{j_p} \exp \{ 2i(\sum_{n=1}^p m_n)~
\theta_l \}~,
\label{summ} \end{equation}
where, $\theta_l ~\equiv~ \pi l /(k+2)$. Expanding the $\sin^2 
\theta_l$ and
interchanging the order of the summations, this becomes
\begin{equation}
N^{\cal P}~=~\sum_{m_1= -j_1}^{j_1} \cdots \sum_{m_p=-j_p}^{j_p}
\left[
~{\bar \delta}_{(\sum_{n=1}^p m_n), 0}~-~\frac12~
{\bar \delta}_{(\sum_{n=1}^p m_n),
1}~-~
\frac12 ~{\bar \delta}_{(\sum_{n=1}^p m_n), -1} ~\right ], \label{exct}
\end{equation}
where, we have used the standard resolution of the periodic Kronecker
deltas in terms of exponentials with period $k+2$,
\begin{equation}
{\bar \delta}_{(\sum_{n=1}^p m_n), m}~=~ \left( 1 \over {k+2} \right)~
\sum_{l=0}^{k+1} \exp
\{2i~[ (\sum_{n=1}^p m_n)~-~m] \theta_l \}~. \label{resol}
\end{equation}

Our interest focuses on the limit of large $k$ and $p$, appropriate to
macroscopic black holes of large area. Observe, first of all, that as $k
\rightarrow \infty$, the periodic Kronecker delta's in (\ref{resol})
reduce to ordinary Kronecker deltas,
\begin{equation}
\lim_{k \rightarrow \infty}~{\bar \delta}_{m_1+m_2+ \cdots +m_p,m}~=~
\delta_{m_1+m_2+ \cdots +m_p,m}~. \label{kinf}   
\end{equation}
In this limit, the quantity $N^{\cal P}$ counts the number of $SU(2)$ singlet
states, rather than $SU(2)_k$ singlets states. For a given set of punctures with
$SU(2)$ representations on them, this number is larger than the corresponding
number for the affine extension. It is here that one makes contact with the
`physical subspace criterion' introduced in Section 3: the $SU(2)$ invariance
is completely natural within this microscopic approach. 

Next, recall that the eigenvalues of the area operator for the horizon,
lying within one Planck area of the classical horizon area $A_H$, are
given by
\begin{equation}
{\hat A}_H~\Psi_S~=~8\pi \beta
~l_{P}^2~\sum_{l=1}^p~[j_l(j_l+1)]^{\frac12}~\Psi_S~,
\label{area} \end{equation}
where, $l_{P}$ is the Planck length, $j_l$ is the spin on the $l$th
puncture on the 2-sphere and $\beta$ is the Barbero-Immirzi parameter
\cite{barb}. We consider a large fixed classical area of the horizon,
and ask what the largest value of number of punctures $p$ should be,
so as to be consistent with (\ref{area}); this is clearly obtained when
the spin at {\it each} puncture assumes its lowest nontrivial value of
1/2, so that, the relevant number of punctures $p$ is given by
\begin{equation}
p~=~{A_H \over 4 l_{P}^2}~{\beta_0 \over \beta}~, \label{pmax}
\end{equation}
where, $\beta_0=1/\pi \sqrt{3}$. We are of course interested in the case
of very large $p$ appropriate to a macroscopic black hole. Observe that 

Now, with the spins at all punctures set to 1/2, the number of states for
this set of punctures ${\cal P}$ is given by
\begin{equation}
N^{{\cal P}}~=~\sum_{m_1= -1/2}^{1/2} \cdots \sum_{m_{p}=-1/2}^{1/2}
\left[ ~\delta_{(\sum_{n=1}^{p} m_n), 0}~-~\frac12~
\delta_{(\sum_{n=1}^{p} m_n),
1}~-~\frac12 ~\delta_{(\sum_{n=1}^{p} m_n), -1} ~\right ] \label{excto}
\end{equation}
The summations can now be easily performed, with the result given {\it precisely}
by the {\it rhs} of eq. (\ref{main}). 

This establishes on a microscopic basis the validity of the extension of the `It
from bit' picture proposed by us in the last section. The central role played by
variables in the doublet representation of a (global) $SU(2)$ group, which we
identified with the binary variables on the lattice approximating the horizon, is
now clarifed. This completes the derivation of our physical space criterion and
the ensuing entropy formula and holographic bound on the basis of a quantum
kinematical formulation. 

We close this section with a remark on the Barbero-Immirzi parameter which has
often led to confusing statements regarding a so-called `ambiguity' within the
quantum geometry approach. The Barbero-Immirzi parameter is an intrinsic aspect of
a Hamiltonian formulation of general relativity based on connection variables
\cite{desa}. Its appearance in the quantum theory is akin to the appearance of the
$\theta$ parameter in quantum Yang-Mills theory in that it signifies the existence
of a one-parameter family of quantum theories corresponding to the same classical
theory. The actual value of $\theta$ can only be decided by experiment. Here in
the case of black hole kinematics, without recourse to experiment, we have
followed \cite{ash} to fix it to match the coefficient of the BHAL which we
believe as the correct lowest order result for macroscopic black holes. This `fix'
works for all non-rotating, generic, four-dimensional black holes. Whether it
works for rotating black holes as well, remains to be seen. But until a
contradiction becomes manifest, there is no reason to suspect that the quantum
kinematics described in this approach has any inherent deficiency. 

\section{Conclusions}

Even though the physical subspace criterion becomes automatic in the quantum
geometry approach, because of special properties of two dimensional WZW theories
at large level and large number of punctures, we would like to emphasize that it
may not be restricted to one particular microscopic construct. Our proposal of
this criterion in Section 3 was made on quite general grounds, and we have shown
that the leading quantum (kinematical) correction follows uniquely from this
criterion, without requiring an underlying microscopic substructure. The existence
of this latter substructure, of course, places the criterion on far firmer grounds
than would have been otherwise possible. However, insofar as macroscopic black
holes are concerned, the full machinery of the quantum geometry approach does not
appear to be exigent. It is another question as to how our criterion by itself can
be used to analyse situations which are qualitatively different in that quantum
{\it dynamical} effects become more important. Our guess is that such situations
require a fuller quantum general relativity than what we have at our disposal at
the moment. 

We should also mention that some recent authors \cite{dasjing} have alleged
that the universality proofs given in \cite{car} of the coefficient of the
$\log (area)$ correction are in fact not technically complete. We have no
comments to offer on this from our heuristic perspective. 

I thank R. Kaul for collaboration leading to understanding of the relevance of
$SU(2)$ in discerning physical horizon states, and for numerous illuminating
discussions.

\end{document}